\documentstyle[titlepage,fleqn,12pt]{article}
\begin{document}

\begin{center}
{\Large \bf{Layer Features of the Lattice Gas Model
\vspace{0.5cm}
for Self-Organized Criticality}}
\vspace{2cm}
{\bf N.C. Pesheva${}^{(1)}$, J.G. Brankov${}^{(1)}$ and
E. Canessa${}^{(2)}$ }
\vspace{0.6cm}
${}^{(1)}$ {\em Institute of Mechanics, Bulgarian Academy of Sciences,}
{\em Acad. G. Bonchev Str 4, 1113 Sofia, Bulgaria}
{\em E-mail:} nina@bgearn.bitnet
${}^{(2)}$ {\em International Centre for Theoretical Physics, Trieste, Italy}
\end{center}
\vspace{2cm}
{\baselineskip=20pt
\begin{center}
{\bf Abstract}
\end{center}
A {\em layer-by-layer} description of the asymmetric lattice gas model
for $1/f$-noise suggested by Jensen [Phys. Rev. Lett. {\bf 64}, 3103 (1990)]
is presented.  The power spectra of the lattice layers in the direction
perpendicular to the particle flux is studied in order to understand how
the white noise at the input boundary evolves, on the average, into $1/f$-noise
for the system.  The effects of high boundary drive and uniform driving
force on the power spectrum of the total number of diffusing particles
are considered.  In the case of nearest-neighbor particle interactions,
high statistics simulation results show that the power spectra of single
lattice layers are characterized by different $\beta_x$ exponents such
that $\beta_x \to 1.9$ as one approaches the outer boundary.

\vspace{2.5cm}
}
\baselineskip=22pt
\parskip=4pt

\pagebreak

Lattice gas models play an important role in exploring self-organized
criticality (SOC) \cite{Bak87} which is an attempt to find a general unifying
principle  explaining the ubiquity of $1/f$-noise and fractality in nature.
The idea behind these studies is that $1/f$-spectra occur when many-body
dissipative systems evolve naturally (without a fine-tuning of some
parameter) towards a set of states without a characteristic length and/or
time scale \cite{Jen90,Voss92,JS92,Can93,Ros93}. The basic feature of
$f^{-\beta}$-noise,
with $\beta \approx 1$, is the long-range and self-similarity of the temporal
fluctuations of some physical quantity such as the total number of diffusing
particles $N(t)$.

In particular, the asymmetric lattice gas (ALG) model introduced by Jensen
\cite{Jen90} allows to describe $1/f$-noise in a highly
interacting system of particles following diffusive dynamics.  The ALG model
can be considered as a transformer which produces a highly correlated signal
out of a white noise (or completely uncorrelated signal) at the input boundary.
It has a simple physical interpretation as a model of transport phenomena in
 type -II superconductors.

In an attempt to understand how the SOC phenomena arise in lattice gas
models, such as the ALG model, it is also necessary to carry out a more
detailed
analysis  in terms of the power spectra of the individual layers in the
model. This is the main goal of this report.

We present here the layer-by-layer description of the Jensen's ALG model
for $1/f$-noise by studying  the power spectra of the lattice layers
in the direction perpendicular to the particle flux.  We studied how the
white noise at the input boundary evolves throughout the lattice, characterized
by different exponents $\beta_{x}$, such that the critical exponent $\beta$
for the total number of particles (and large times) is close to 1 as
first found by Jensen \cite{Jen90}.  The effects of high boundary drive
and uniform driving force on the power spectrum of the total number of
diffusing particles are also considered.

The lattice gas cellular automaton model of Jensen \cite{Jen90} is a
system of particles residing on a square lattice $\Lambda$ of
$N_x\times N_y$ lattice sites. The particles obey an exclusion principle, so
that there is no more than one particle
on a given site at time $t$. With every site $\vec r\in \Lambda$
a variable $n(\vec r;t)$ taking values $0,1$ is associated.  As usual
$n(\vec r;t)=1$ means that there is a particle on site $\vec r$ at time $t$
and $n(\vec r;t)=0$ that the site is empty. The particles on nearest neighbor
sites are repelling each other with a central force of strength $J$,
{\em i.e.}
$$
 \vec F_{int}(n(\vec r;t))=-Jn(\vec r;t)\sum_{i=1}^q n({\vec r+ \vec e_i};t)
 \vec e_i \ \ ,\eqno(1)
$$
where $ \vec e_i\ , i=1,\dots,q$ are the unit vectors to the nearest neighbors
(in the calculations we put $J=1$).
In the more general case an additional driving
force $\vec F_{dr}(\vec r)$ could be applied to the particles
so that the total force acting on a particle is
$$
 \vec F_{tot}(\vec r)=  \vec F_{int}(\vec r) + \vec F_{dr}(\vec r) \ .\eqno(2)
$$
The system evolves at discrete time steps following diffusive deterministic
dynamics. The configuration ${\{n(\vec r;t)\} }_{\vec r  \in \Lambda}$ of the
system is updated simultaneously. Every particle $l$
is moved one lattice site in the direction of the resulting force, so that
the new coordinates of the particle are
$$
\vec r_l(t+1)= \vec r_l(t) + \vec d_l(t) \   , \eqno(3)
$$
where $\vec d$  is the displacement vector, determined by
$$
d_x  =\left[ F_x \over F \right] \;\;\;  ;  \;\;\;
d_y  =\left[ F_y \over F \right] . \eqno(4)
$$
Here $ F=| \vec F| $ and
the square brackets $[.]$ mean taking the nearest integer to the number
enclosed in the brackets. Equation (3) in combination with (4) implies that
diagonal moves are also allowed.  \par
 To ensure single occupancy of a lattice site, some additional rules,
 termed by Jensen \cite{Jen90} blocking mechanisms, are applied.
 These are:
\begin{description}
\item [1)]  If the site to which a particle attempts to move is occupied, the
     particle is not moved;
\item [2)] If there are two particles which attempt to move to the same
lattice site, the particle to which a larger force is applied wins ; if
the forces are equal - neither particle moves.
\end{description}

The boundary conditions are asymmetric.  It is assumed that at the left side
of the system ({\em i.e.} at $(0,y)$) there is a layer of
fixed particles . The role of this layer is to push the
particles in the first column $(1,y)$ into the system. At every time step
the particles in the first column are first removed and then new particles
are introduced with a probability $p$, called boundary drive.
In this work, the effect of the lattice boundary drive $p$ is studied over a
large range of possible values (0-1).  The particles can freely leave the
system over the right edge. Periodic boundary conditions are imposed
in $y$-direction. This leads to a net flux of particles through the system.

  The power spectrum $S_N (f)$ of the total number of particles
in the system is defined in a standard way as \cite{Ben71,Bon94}
$$
S_N(f)=\lim_{T\to\infty} {2\over T} {\left<| X(f,T)|^2 \right>} \ , \eqno(5)
$$
where  $X(f,T)$ is the finite Fourier transform of $N(t)$
(from sample sequencies $N(t_i)\ ,\ i=1,\dots , T$) given by
$$
X(f,T)=\int_0^T N(t) e^{-i2\pi ft} dt \eqno(6)
$$
and the angular brackets $<>$ denote averaging over many realizations of the
process.

If the process is stationary  $S_N(f)$ can be expressed in terms of the
autocorrelation
function of the time signal $R_N(\tau) =\lim _{T\to\infty} {1\over T} \int_0^T
\left< N(t)N(t+\tau)\right> dt$ (the so-called Wiener-Khintchin's theorem)
$$
S_N(f)=4\int_0^{\infty} R_N(\tau) \cos \left( 2\pi f\tau \right)  d\tau  \ .
\eqno(7)
$$ \par

We focus now on the results obtained for the system from the present
simulations.
 Our simulations  confirm that the power spectrum has a power low
dependence on the frequency in a wide interval of frequencies, {\em i.e.}
$S_N (f) \sim 1/f^{\beta}$ , with $\beta$ roughly equal to 1.  As can be seen
from Fig.1, at small frequencies $f<f_c$,  there is a crossover
to a white noise ($S_N(f)\sim \ {const.}$) due to the finite system size.
Indeed, $f_c$ decreases with increasing the system size as $\sim 1/{N_x}^2$
as predicted in \cite{Fii93} but with different proportionality constant.

Having in mind that when the system size $N_x\to \infty$  the power spectrum
is expected to be $\sim 1/f^\beta$, we approximate the power spectrum for the
finite system by
$$ F(f)= \frac {A}{ {\left( 1+{\left( f/f_c \right)}^2 \right)}^{\beta/2} }
 \quad , \eqno(8)
$$
where $A$, $f_c$ and $\beta$ are fitting parameters.  We  find that the
best fit to our results for several system sizes $N_x \times 8$,
$N_x=10,20,30,40,50,100$ for $A$, $f_c$ and $\beta$ gives:

$$
A\approx 0.604 {f_c}^{-\beta},\  {\beta\approx 1.09}  \eqno(9)
$$
as one would expect from Eq.(8).

When we change the boundary drive $p$ between 0 and 1, $\beta$ is found to
vary slightly.  However, the limits $p\to 0$ and $p\to 1$ need special
attention.  For the case $p\to 0$, as studied in Refs. \cite{Jen90,Fii93},
it was reported a transition towards $\beta\approx 2$.  Whereas for a high
boundary drive, {\em e.g.} $p=0.99$, we have found $\beta \approx 1.4$ as
compared to $\beta\approx 1.05$ when $p=0.90$ for the same system size.  Our
simulation results also show that for a high boundary drive there exists a
tendency of $\beta$ to grow as the system size is increased. For $p=0.9$ and
size $60\times 48$ we find $\beta\approx 1.4$ as compared to $\beta\approx
1.05$
for the small $10\times 8$ system.

In the limiting case $p=1$ the system is completely deterministic and can
be studied analytically.  The power spectrum is changed to a new type,
consisting only of a few characteristic frequencies.  For a system of
length $N_x=8$, the spectrum consists of the following set of frequencies
$\{0,\ 0.125,\ 0.250,\ 0.375,\ 0.5\}$.  The presence of
even a very small stochastic element in the boundary drive, {\em i.e.}
$p\to 1$, but $p \ne 1$ produces a qualitatively different picture.

Let us continue studying now  the individual
layers in the asymmetric lattice gas model under consideration.
The averaged occupancy of a layer of the lattice $<\overline{n(x)}>$ is
defined as
$$
<\overline{n(x)}>=\left< {1\over T}\sum_{i=1}^{T} \sum_{y=1}^{Ny} n(x,y,t_i)
\right>\ ,
\quad  x=1,\dots,N_x \ .
\eqno(10)
$$
At low boundary drive $p=0.2$, we obtain a typical `saw-like' shape of the
density profile for small lattice sizes.  At this value of $p$ the average
density of particles in the statistically stationary state of the system
remains low, $\rho \approx 0.28$.  Since the particles are repelling each
other,
states with alternating high occupancy layers and low occupancy layers are
encountered more often.  As the system size is increased the density
distribution gets more smooth as shown in Fig.2, but it can not be well
approximated by the linear distribution with constant gradient as discussed
in \cite{Jen90}.  An increase in the boundary drive smoothens the density
distribution.  We add that the application of a driving force lowers the
average
occupancy in the system as compared to the case without driving force.

At low boundary drive, the application of a small uniform driving force
destroys the $1/f^{\beta}$ type of the power specrtum as seen in Fig.3.
For $F_{dr} > F_{cr}$, where  $F_{cr}\approx 1.75$, the power spectrum is
the same as for a noninteracting system (except for the hard-core exclusion)
as shown in Fig.4.  At higher values of the boundary drive and small driving
force the power spectrum displays again the $1/f^{\beta}$ behaviour in a
small frequency interval.  Therein $\beta$ takes a value ($\beta \approx 1.6 $)
larger than in the case without driving force ($\beta \approx 1.05$), as shown
in Fig.5. This behaviour could possibly be explained in the following way. The
higher boundary drive ($p=0.9$) produces higher average
density in the system. A higher density leads to stronger correlations and
damps the effect of the driving force. This point needs further elucidation
in view of the analytical results obtained from Langevin equation \cite{GHJ92}.

At a moderate boundary drive, $p=0.2$, the autocorrelation functions and the
individual power spectra of the layers are studied for different lattice sizes,
$N_{x} \times 8$, where $N_{x}=10,20,30,50,100$ and also $60\times 48$.
The power specrtum of the number of particles in a layer, $S_{L_x}(f)$, has
a form similar to that for the total number of particles in the system as seen
in Fig.6.
The critical frequency $f_c$ at which a crossover to white noise occurs depends
on the layer selected, {\em i.e.} $f_c=f_c(x)$.  The width of the frequency
interval $\Delta f$, in which $1/f^{\beta}$ holds, depends also on the position
$x$.
As $x \to N_x$, the crossover frequency changes toward higher frequency values,
and the interval $\Delta f(x)$ broadens.  It is interesting to note that even
the first layer (immediately after the input boundary layer - whose power
spectrum corresponds to white noise) already has $\beta\approx 0.8$,
though in a very narrow frequency interval.

We have calculated the autocorrelation functions $R_{L_{x}}(\tau)$ of the
system
layers (shown in Fig.7)
by averaging $N_{L_{x}}(t)N_{L_{x}}(t+\tau)$ over one simulation and
then averaging over many (up to $1000$) independent simulations for
different system sizes.  $R_{L_{x}}(\tau)$ of the single layers are found to
decay much slower than the autocorrelation function of the whole system.
As discussed in Ref. \cite{Yor94}, finite system effects on $R_{L_{x}}(\tau)$
should be taken into account to get the proper behaviour for the infinite
system.

In Fig.8, $\beta_{r}$ data for several system sizes are shown as a function
of the normalized  layer distance within the lattice $r=x/N_x$.
We find that the exponent of the last layer is $\beta(r=1)\approx 1.9$
-- a result very close to the well known result for the Wiener process.
Furthermore we also see that, within numerical accuracy (less than
$\sim 5\%$ error), all $\beta_{r}$ lie on approximately the same curve.
Thus, we  find that while the individual power spectra of the layers are
different and characterized with different $\beta_{x}$, their collective
behavior produces a power spectrum
for the total number of particles in the system with $\beta \approx 1.1$
(Fig.1).

We also studied the effect of applying a nonuniform driving force on the power
spectrum of the total number of particles in the ALG. For moderate boundary
drive ($p=0.2$)  we applied additional driving force with a constant gradient
$\Delta F_{dr}=0.1$. We find that the effect on the power spectrum is much less
pronounced
than in the case of a uniform driving force. In fact one can still approximate
the spectrum by $1/f^\beta$, but now $\beta\approx 1.2$ for a system of size
$10\times 8$ as compared to $\beta\approx 1.05$ for the same system without
shear.

In conclusion we summarize our numerical findings.  In an attempt to understand
how SOC arises in simple lattice gas models of many particle systems, we have
carried out a more detailed
analysis of the   ALG model in terms of the power spectra of the individual
layers in direction perpendicular to the particle flux. The effects of high
boundary
drive and uniform driving force on the power spectrum of the total number of
diffusing particles have been also considered.  In particular, our results
for the limiting case $p\to 1$ supplement previous work \cite{Jen90}.
Our findings  reveal some interesting features for the spatial variation
of the (local) exponents $\beta_{x}$.  The layer after the boundary
layer with the white noise power spectrum, is found to display an exponent
$\beta(x=2) \approx 0.8$ in a very narrow frequency interval.
Whereas, we find that the exponent of the last layer is $\beta(r=1) \approx
1.9$.

\begin{center}
{\bf  Acknowledgments}
\end{center}

N.C.P. gratefully acknowledges the EEC fellowship -- GO WEST hosted by the
ICTP, Trieste, where this work was started.

This work was supported by the Bulgarian National Foundation for Scientific
Research Grant No. MM-405/94.

\newpage
\onecolumn

\section*{Figure captions}

\begin{itemize}
\item {\bf Fig.1}:  Log-log plot of the power spectra $S_N (f)$ of the
total number
of particles against frequency $f$ at low boundary drive $p=0.2$ and no driving
force:  curve (1) is for a system of size $10\times 8$, curve (2) is for a
system of size $50\times 8$ together with the continuous fitting function
$F(f)$.

\item {\bf Fig.2}:  The average occupancy of a column $<n(x)>$ versus column
number for a system $100\times 8$ - curve (1) and for a system $50 \times 8$
- curve (2) at $p=0.2$.

\item {\bf Fig.3}:  The effect of a small uniform driving force $F_{dr}=0.1$
on the power spectrum for a system $10\times 8$ at small boundary drive
$p=0.2$.

\item {\bf Fig.4}:  The effect of a small uniform driving force $F_{dr}=0.1$
on the power spectrum for a system $10\times 8$ of a noninteracting system
(hard-core interaction only).

\item {\bf Fig.5}:  Log-log plot of the power spectra of a system $10\times 8$
at high boundary drive $p=0.9$: curve (1) for $F_{dr}=0.1$ ($\beta \approx
1.6$)
and curve (2) for $F_{dr}=0$ ($\beta \approx 1.05$).

\item {\bf Fig.6}:  Log-log plot of the power spectra of two layers of a
system of size $30\times 8$ are shown : curve (1) is the power spectrum
of the middle layer -- ${x}=15$ ($\beta \sim 0.95$) and curve (2) of the
last layer -- ${x}=30$ ($\beta \sim 1.9$).

\item {\bf Fig.7}:  The normalized autocorrelation functions $R_{L_{x}}(\tau)$
corresponding to the power spectra in Fig.6: curve (1) is the autocorrelation
function of the middle layer ${x}=15$  and curve (2) is the autocorrelation
function of the last layer ${x}=30$.

\item {\bf Fig.8}:  The dependence of the critical exponent ${\beta}_r$ on
the relative distance of a layer number $r$ at moderate boundary drive
$p=0.2$ for system sizes $N_{x}\times 8$ with $ N_{x}=20,30,40,50,100$.
Curve (1) is for a $20 \times 8$ system and curve (2) is for a $100 \times 8$
system.

\end{itemize}

\end{document}